# Scalable van der Waals epitaxy of tunable moiré heterostructures

Matthieu Fortin-Deschênes[1*], Kenji Watanabe[2], Takashi Taniguchi[3], Fengnian Xia[1*]

[1]Department of Electrical Engineering, Yale University, New Haven, Connecticut 06511, United States

[2]Research Center for Functional Materials, National Institute for Materials Science, 1-1 Namiki, Tsukuba 305-0044, Japan

[3]International Center for Materials Nanoarchitectonics, National Institute for Materials Science, 1-1 Namiki, Tsukuba 305-0044, Japan

**The unique physics found in moiré superlattices of twisted or lattice-mismatched atomic layers hold great promise for future quantum technologies[1,2,3,4]. However, twisted configurations are typically thermodynamically unfavorable, making the accurate twist angle control in direct growth implausible[5,6]. While rotationally aligned moiré superlattices based on lattice-mismatched layers such as $WSe_2/WS_2$ can be synthesized[7,8,9,10,11,12,13,14,15,16,17,18,19], they lack the critical tunability of the moiré period and the moiré formation mechanisms are not well-understood. Here, we report the scalable, thermodynamically driven van der Waals epitaxy of stable moirés with tunable period from 10 to 45 nanometers, based on lattice mismatch engineering in two WSSe layers with adjustable chalcogens ratios. Contrarily to conventional epitaxy, where lattice mismatch induced stress hinders high-quality growth, we reveal the key role of bulk stress in moiré formation, as well as its unique interplay with edge stress in shaping the moiré growth modes. Moreover, the synthesized superlattices display tunable interlayer, and moiré intralayer excitons. Our studies unveil the unique epitaxial science of moiré synthesis and lay the foundations for moiré-based technologies.**

Email: matthieu.fortin-deschenes@yale.edu, fengnian.xia@yale.edu

**Introduction**

Overlaying the lattices of two-dimensional (2D) materials with a relative twist angle or a lattice mismatch generates a larger (quasi-) periodic moiré pattern[20,21,22,23]. The moiré drastically influences the properties of van der Waals heterostructures (vdW HSs) due to the formation of a moiré potential with long periodicity and the emergence of additional moiré electronic bands. In moiré transition metal dichalcogenides (TMDCs), there exist intralayer and interlayer exciton minibands as well as moiré trapped excitons[1,24,25,26,27,28]. In "magic-angle" twisted bilayer graphene (TBG), the minimized electron kinetic energy in the moiré bands leads to strongly correlated phases such as unconventional superconductivity and ferromagnetism[29,30,31,32]. Experimentally, moiré HSs are usually fabricated by stacking mechanically exfoliated layers with a precisely controlled twist angle[22] or by rotationally aligning lattice-mismatched layers[24,25,28]. The accurate rotational misalignment allows to finely tune the properties of the HSs. Nonetheless, this method suffers from important limitations such as the introduction of interlayer contamination. More importantly, mechanical exfoliation is inherently not scalable, which hinders the integration of moiré materials in emerging quantum technologies.

Large-scale synthesis can overcome the fundamental limitations of mechanical exfoliation[33]. While the direct synthesis of twisted moiré graphene has been demonstrated by vdW epitaxy[5,6], the accurate control of the rotational misalignment is very challenging, since twist angles away from main crystallographic directions are almost all equally thermodynamically unfavorable. Furthermore, TBG was recently synthesized by etching Cu/TBG/Cu sandwich structures obtained from the growth of monolayer graphene on two pre-rotated single crystalline Cu foils[34]. Rotationally aligned moirés such as graphene/hBN and various TMDCs have also been grown by vdW epitaxy[7,8,9,10,11,12,13,14,15,16,17,18,19]. Despite these progresses, foundational issues for moiré vdW epitaxy, such as the mechanisms and conditions for the moiré formation, have not yet been addressed. More importantly, in these earlier studies, the period of rotationally aligned moirés lacks the critical tunability and is restricted to a handful of discrete values determined by the lattice parameters of the two layers. For these reasons, the synthesis of two aligned atomic layers with controllable lattice mismatch represents a novel strategy for realizing large-scale stable moirés with tunable periodicity (Fig. 1a). In fact, in most 2D materials, interchanging p-block elements from the same group of the periodic table allows to change the lattice constant while minimally modifying the electronic band structure.

Herein, using TMDCs as a model system, we demonstrate that substitutional alloying of p-block elements allows to continuously tune the moiré period of vdW HSs using a thermodynamically driven direct growth process. We grow $WS_ySe_{2-y}/WS_xSe_{2-x}$ vdW HSs with different sulfur/selenium ratios using chemical vapor deposition (CVD) and obtain moiré patterns with tunable period between 10-45 nm. We then elucidate the moiré formation mechanisms and reveal the unique growth modes based on transmission electron microscopy measurements and modelling of moiré nuclei. Finally, by examining the photoluminescence (PL) of the vdW HSs, we demonstrate the tunability of the interlayer exciton as well as the presence of intralayer moiré excitons. Our work identifies the key mechanisms driving moiré formation and provides a scalable approach for the integration of tunable, thermodynamically stable moiré semiconductor materials in emerging quantum technologies.

**Growth of van der Waals heterostructures with tunable moiré**

Moiré HSs were grown on 300 nm $SiO_2$/Si substrates by CVD using a two-step process, as illustrated in Fig. 1b and 1c. First, sulfur-rich monolayer $WS_xSe_{2-x}$ was grown using S, Se and $WO_{2.9}$/NaCl precursors, with argon carrier gas, as shown in Fig. 1b (details in Methods). The grown $WS_xSe_{2-x}$ mostly consists of monolayer triangular flakes with lateral dimensions between 50-200 μm, as shown in Fig. 1b and Fig. S1. In the second step, selenium-rich $WS_ySe_{2-y}$ monolayer was grown on $WS_xSe_{2-x}$ using a mixture of argon and hydrogen carrier gas and a larger Se/S ratio (Fig. 1c). $WS_ySe_{2-y}$ forms a continuous monolayer on the $WS_xSe_{2-x}$ flakes, creating $WS_ySe_{2-y}$/$WS_xSe_{2-x}$ vertical vdW HSs, as seen in the optical micrograph in Fig. 1c. We note that multilayer $WS_ySe_{2-y}$ grows on $SiO_2$ (teal layer in Fig. 1c), as well as at the edges of the flakes, and on the central nuclei of the $WS_xSe_{2-x}$ triangular flakes (Fig. 1c). Moreover, $WS_ySe_{2-y}$ grows in a quasi layer-by-layer fashion on monolayer $WS_xSe_{2-x}$ (Fig. S2). Additional partial growth experiments suggest that the $WS_ySe_{2-y}$ layer is formed by coalesced, rotationally aligned, (2H and 3R) domains with lateral dimensions up to 10 μm (Fig. S2). While not studied in detail in this work, controlling the nucleation may allow to obtain pure 2H domains by vdW epitaxy (Fig. S3) [11,35,36,37,38,39].

Next, we look at the atomic structure of the vdW HSs using transmission electron microscopy (TEM). A TEM micrograph of the moiré unit cell of the $WS_{0.35}Se_{1.65}$/$WS_{1.88}Se_{0.12}$ sample shown in Fig. 2a can be seen in Fig. 2b. The two layers are rotationally aligned, but have different lattice parameters (3.16 Å and 3.26 Å), as evidenced by the two sets of spots with hexagonal symmetry in the Fourier transform (FFT) of the large-scale TEM image (yellow squares in Fig. 2c). Moreover, the intensity of the two sets of spots, both in the FFT (Fig. 2c) and diffraction (Fig. S4), is comparable. This indicates that the two layers have the same thickness (i.e. monolayers), in agreement with the optical microscopy contrast (Fig. 2a). The smaller hexagonal pattern (magenta squares in Fig. 2c)) is associated to a large-scale moiré with a 10.4 nm lattice constant. Furthermore, the reciprocal lattice vectors of the moiré and atomic lattices are aligned within 1°, as measured from the FFT in Fig. 2c. This implies that the two layer are aligned within ~0.03°, as will be discussed in the next sections.

The effectiveness of moiré compositional tuning is assessed by comparing vdW HSs grown under different Se/S ratios. The composition of the HSs is determined using energy-dispersive X-ray spectroscopy (EDX) as shown in Fig. S5 (details in methods). The selenium content difference between the two $WS_ySe_{2-y}/WS_xSe_{2-x}$ layers, defined as $\Delta Se = (x - y) / 2 \times 100\%$, ranges from 17-77% and moiré patterns are formed in every sample. The moiré period increases from 10 to 45 nm when $\Delta Se$ decreases from 77% to 17% as illustrated in Fig. 2d, due to the narrowing equilibrium lattice mismatch. The expected relaxed moiré lattice parameter is:

$$a_{moiré} = \frac{a_{WS_xSe_{2-x}} a_{WS_ySe_{2-y}}}{\left(a_{WS_ySe_{2-y}} - a_{WS_xSe_{2-x}}\right)} \approx \frac{a_{WSSe}^2}{\Delta Se(a_{WSe_2} - a_{WS_2})} \quad (1)$$

where $a_{WS_xSe_{2-x}}$ is the lattice constant of monolayer $WS_xSe_{2-x}$. The measurements agree with the calculated moiré periodicity based on equation (1) as illustrated in Fig. 2e, indicating that the two layers are relaxed to their equilibrium lattice constant. However, for small lattice mismatch ($\Delta Se = 17\%$), partial relaxation and larger moiré periodicities are observed in some regions, as discussed in the next sections.

**Stress relaxation and moiré formation mechanisms**

To better understand moiré formation, we look at the nucleation and stress relaxation mechanisms. Below, we consider the energy landscape of small rotationally aligned $WS_ySe_{2-y}$ nuclei on continuous $WS_xSe_{2-x}$. The bottom $WS_xSe_{2-x}$ layer is assumed to remain relaxed since its continuity and interaction with the growth substrate restrict local deformations. The energy difference per unit area between relaxed nuclei and lattice-matched (epitaxial) nuclei ($\Delta E$) can be understood as:

$$\Delta E = \Delta E_{int} + \Delta U_{bulk} + \Delta U_{edges} \qquad (2)$$

Here, the first two terms on the RHS are the interlayer interaction energy cost of leaving the epitaxial configuration and the bulk strain energy released during relaxation. The edge strain energy term ($\Delta U_{edges}$) originates from the modified equilibrium interatomic distances at the edges of the nuclei, due to dangling bonds. The edge stress ($\tau_{edges}$) is expected to be tensile for TMDCs[41].

We highlight the importance of the three terms in Equation (2) by estimating the energy of triangular nuclei as a function of their lattice constants. To keep our analysis general to vdW HSs, we use a simple model based on a 2D hexagonal lattice (details in Fig. S6 and Methods). Since the two layers are coupled only by weak vdW interactions, we compute the intralayer and interlayer interactions independently. The intralayer interactions are modelled by harmonic bonds with force constants fitted to experimental data[42], and edge stress is introduced by shortening the equilibrium bond length at the edges of the nuclei. On the other hand, the interlayer interactions are modelled by a 2D periodic potential with a well at each lattice site, as detailed in the Methods section. Here, $WS_2$ is used as the bottom layer in the calculations. The three contributions to the energy can then be directly separated and are plotted as a function of the lattice parameter of the top layer in Fig.

3a and 3b for nuclei with side length ($L$) of 8 nm and 56 nm, respectively ($\Delta$Se=25%). The bulk strain energy ($U_{bulk}$) favors the relaxed configuration, and the interlayer interaction energy ($E_{int}$) favors the lattice-matched epitaxial state (Fig. 3a and 3b). However, $U_{bulk}$ is independent from $L$, but $E_{int}$ narrows as $L$ increases since the average atomic displacements from the lattice-matched sites, resulting from the nuclei's expansion or contraction, scales with $L$. On the other hand, the tensile edge strain energy per unit area ($U_{edges}$) scales as $1/L$ and favors the compression of the nuclei (Fig. 3a and 3b). The relative magnitudes and signs of these three energy terms have a fundamental role in shaping the moiré growth modes, as discussed below.

The evolution of the energy *vs*. lattice constant for WS$_y$Se$_{2-y}$ nuclei of increasing $L$ is shown in Fig. 3c and 3d for $\Delta$Se=25% and $\Delta$Se=30%, respectively. Due to the broad nature of $E_{int}$ in small nuclei, the $E_{int}$ and $U_{bulk}$ minima merge into a single one for small $L$, as illustrated in Fig. 3c and 3d. The path followed by the nuclei to minimize their energy as the two minima separate during lateral growth (black arrows in Fig. 3c and 3d), dictates the resulting moiré growth modes, which are illustrated in Fig. 3e to 3g. When $-\frac{\Delta U_{bulk}}{\Delta E_{int}} < 1$, the lattice-matched configuration is favored and epitaxial growth of strained vdW HSs is expected, as illustrated by the striped region in Fig. 3e. On the other hand, if the lattice mismatch is sufficient, aligned moirés become the thermodynamically stable configuration (middle region of Fig. 3e). The magnitude and sign of the edge and bulk stress can be expected to lead to two distinct aligned moiré growth modes, which are illustrated in Fig. 3f and 3g. If the lattice mismatch is relatively small, such as for $\Delta$Se=25% in Fig. 3c, nuclei get trapped in the lattice-matched epitaxial configuration, which eventually becomes a metastable state as the top layer nuclei grow. Nuclei can then transition to the stable relaxed configuration (dashed arrow in Fig. 3c) and thus form a stable moiré, which expands by lateral growth, as illustrated in Fig. 3f. The trapping in the metastable state is mostly driven by edge stress, which prevents relaxation if it has an opposite sign to the bulk stress. However, since $U_{edges}$ scales as $1/L$, the relaxed state eventually becomes the most stable state. The conditions favoring this growth mode are summarized by the yellow region in Fig. 3e. We note that the metastable state can form even without edge stress, due to the narrowing of $E_{int}$, as detailed in Fig. S7. The second growth mode (green region in Fig. 3e) arises in conditions where the lattice-matched metastable state is never favored, such as in Fig. 3d. This occurs at larger equilibrium lattice mismatch, or when the edge stress is small or has the same sign as the bulk stress (which can be achieved by reversing the growth order in this work, for instance). These conditions can be expected to lead to direct moiré growth, meaning that a relaxed moiré grows laterally immediately after nucleation, as shown in Fig. 3g. A large-scale moiré likely formed by the direct moiré growth mode, given the large lattice mismatch (ΔSe=77%), is presented in Fig. 3h. Furthermore, as the lattice mismatch increases even further, random rotational misalignment of the two layers is eventually expected, as happens in many materials systems[43] (right side of Fig. 3e).

TEM images of the early stages of the two aligned moiré growth modes are displayed in Fig. 3i and 3j for ΔSe=38% and 17%, respectively. At ΔSe=38%, the lattice mismatch is large enough to lead to direct moiré growth. In fact, small nuclei (150 nm) already display a moiré pattern (Fig. 3i). On the other hand, a fundamentally different behavior, associated to the other moiré growth mode, is observed for ΔSe=17%. Fig. 3j shows a large ~1 μm nucleus (ΔSe=17%) ongoing stress relaxation, which confirms the presence of a metastable lattice-matched state in vdW HSs with small lattice mismatch. In fact, no moiré is observed on most of the nucleus area, indicating that despite the chalcogens ratio difference between the top and bottom layers, the epitaxial configuration is favored during nucleation. Interestingly, this intermediate state provides deeper insight into the stress relaxation mechanisms. Given the nucleus size and corrugation of the interlayer interaction energy[44], the energy barrier for concerted relaxation can reach tens of keV. This means that relaxed regions must nucleate first and then grow. Relaxation cannot nucleate in the bulk since it would be surrounded by an energetically prohibitive high-stress region. Hence, stress relaxation is expected to nucleate at the edges or corner of the nuclei and spread inside the material. The TEM data confirms that stress relaxation nucleates at the nuclei's corners, as indicated by the parallel stripes (1D moiré) in Fig. 3j. The orientation of the stripes implies that corners are first pushed out to relieve stress uniaxially, parallel to the corner's bisector. Indeed, the periodicity of the outmost stripes is ~35 nm, meaning that the stress is fully relieved in one direction at the corners. On the other hand, the nucleus is still strained in the other direction, suggesting that the edges of the triangle are pinned to the bottom layer. Nonetheless, biaxial stress relaxation (stripes at ~60° from the parallel stripes) is seen at the edges of the triangular nucleus (red circle in Fig. 3j) and seems to occur in a later stage. Furthermore, we carried out molecular dynamics simulations of stress relaxation in strained metastable nuclei, which confirm that relaxed regions first nucleate at corners and spread inside the material (Fig. S8).

**Moiré homogeneity**

The moiré homogeneity is closely related to the growth and stress relaxation mechanisms. We analyzed the strain and twist angle homogeneity by fitting the moiré lattice vectors in TEM images such as the one displayed in Fig. 4a for $\Delta Se=53\%$. Using the average reciprocal lattice vectors of the bottom layer, measured in the FFT, we then calculated the top layer lattice vectors for every moiré unit cell with $\vec{k}_{moiré} = \vec{k}_{WS_xSe_{2-x}} - \vec{k}_{WS_ySe_{2-y}}$. The average twist angles (θ) of the analyzed samples (ΔSe between 17-77%) are <0.037° and local fluctuations of 0.02-0.06° (standard deviation) are observed in the analyzed regions (~250 × 250 nm²), as illustrated in Fig. 4b for ΔSe=53%. Fig. 4c shows the local inhomogeneities of the lattice mismatch, which is on the order of 0.018-0.0028 Å (SD). The twist angle distribution for ΔSe=53% is shown in Fig. 4d, and Fig. 4e presents the distributions of the lattice mismatch for three different compositions. These small inhomogeneities of the strain and twist angle (Fig. 4 (b-e)) may form during lateral growth of the top layer or originate from strain heterogeneities created during cooldown of the first layer. In fact, partial stress relaxation can happen for monolayer TMDCs on $SiO_2$ when cooling down from 850 °C [45,46]. This suggests that using a process that allows to keep the temperature constant between the growth of the two layers or growing on substrates such as sapphire where relaxation occurs during cooldown[47], could improve the homogeneity even further.

While the moirés of vdW HSs with substantial lattice mismatch are typically homogeneous, irregular moirés with long periodicity can form when the lattice mismatch is very small, especially in smaller flakes. In fact, when the interlayer interactions and strain energy are comparable (e.g., ΔSe=17%), nuclei are trapped in the metastable lattice-matched configuration. Since relaxation occurs in larger nuclei and is slower, this can lead to the presence of residual strain such as in Fig. 4f. In fact, the relaxation is not rapid enough to be completed before a substantial area is grown. Both relaxed regions (such as in Fig. 2d) and partially strained regions (Fig. 3j and Fig. 4f) are observed at ΔSe=17%, which suggests that this composition approaches to the lower limit for moiré formation.

Inverting the two growth steps can be considered as a strategy to promote stress relaxation and improve homogeneity in samples with very small lattice mismatch. In fact, the edge stress is expected to be tensile in TMDCs[41] and therefore competes with the compressive bulk stress of the top Se-rich layer. This favors the metastable lattice-matched state and increases the energy barrier for relaxation. On the other hand, if the S-rich layer is grown on top of the Se-rich layer, the bulk stress of the nuclei become tensile. The tensile edge stress would therefore act as an additional driving force for relaxation, which would lead to early stress relaxation or even promote the direct moiré growth mode. Moreover, this could potentially allow to realize even larger moiré periodicities by allowing stress relaxation in vdW HSs with smaller lattice mismatch. Other strategies can also be considered to improve the homogeneity of small lattice mismatch moirés, including using higher temperature and lower growth rate, or implementing a two-step growth process to have favorable stress relaxation followed by lateral growth of the desired composition.

**Excitons in grown moiré heterostructures**

Next, we look at the photoluminescence (PL) of $WS_ySe_{2-y}/WS_xSe_{2-x}$. To improve the emission intensity, the grown HSs were removed from the substrate and then encapsulated with hBN for optical measurements (Fig. 5a)[48]. The PL (532 nm excitation, T=77K) is dominated by interlayer exciton (IX) recombination due to the rapid charge transfer at the vdW interface[49] (Fig. 5b). $WS_ySe_{2-y}/WS_xSe_{2-x}$ can be expected to form type II heterostructures for all compositions[50] and therefore electrons (holes) generated in the sulfur- (selenium-) rich layer transfer to the selenium- (sulfur-) rich layer to form the IX (Fig. 5b). Expectedly, the IX peak position (1.44 eV) for $\Delta Se=77\%$ is slightly blue-shifted from previous results on $WSe_2/WS_2$ (1.41 eV)[25]. Furthermore, as ΔSe decreases to 36%, the IX PL peak blue shifts from 1.44 eV to 1.59 eV due to the shrinking band offset.

In as-grown monolayer $WS_xSe_{2-x}$, a single peak associated to the A exciton is observed and blue shifts with increasing S content (Fig. 5c)[51]. In the vdW HSs, PL from the intralayer A excitons of both layers can be observed, as shown in Fig. 4d. The intralayer exciton emission intensity is ~100 times weaker than the IX one (Fig. 5b). Furthermore, the signal of the Se-rich layer is larger than the signal of the S-rich layer. Like in minimally twisted $WSe_2/WS_2$ and $MoSe_2/MoS_2$, the A exciton of the Se-rich layer splits into three peaks in the vdW HSs, due to the formation of exciton minibands in the moiré potential[25,28]. For ΔSe=77%, the peaks are located at 1.74, 1.82 and 1.89 eV (brown curve in Fig. 5d), slightly blue-shifted from previous measurements on mechanically exfoliated and stacked $WSe_2/WS_2$[25]. Nonetheless, the 150 meV splitting is larger than in minimally twisted $WSe_2/WS_2$ (~100 meV), which may be due to stronger interlayer coupling in grown samples. Expectedly, as the top layer S content increases, the moiré intralayer exciton peaks blue shift. We also note some spatial variation of the peak positions and intensities for the intralayer moiré excitons, which may be associated with inhomogeneities introduced during the transfer process. Nonetheless, these results demonstrate that intralayer moiré excitons form in grown vdW HSs.

**Summary and Outlook**

In summary, we demonstrated that substitutional alloying of S/Se allows for the thermodynamically driven growth of TMDC vdW HSs with tunable moiré periods. Furthermore, we identified the critical role of bulk and edge stress in shaping the conditions favorable to moiré formation and the resulting moiré growth modes. At large lattice mismatch, large-scale relaxed moirés grow directly after aligned nucleation. At small lattice mismatch and when the bulk and edge stress have opposite signs, metastable lattice-matched nuclei first form and can reach lateral dimensions up to a micrometer. Stress relaxation then nucleates at the corners of the nuclei and spreads into the bulk, which then allows for lateral growth of relaxed moirés. This new understanding can be harnessed to control the large-scale synthesis and engineer novel moiré nanostructures, representing a critical step for the integration of moiré vdW HSs in emerging quantum technologies.

Combining this approach with nucleation engineering and kinetics control to obtain pure 2H or 3R stacking (Fig. S3) will allow to extend moiré epitaxy to the wafer-scale[11,35,36,37,38,39]. Furthermore, alloying can be extended to tune other moiré materials containing p-block elements, such as other TMDCs, 2D pnictogen alloys[52], magnetic metal halides[53], etc. In light of the recent progress made on the synthesis of 2D silicon-carbon alloys[54], graphene/2D-$C_xSi_{1-x}$ vdW HSs with only ~3.5% silicon[55] can be targeted to reproduce the properties of magic angle twisted bilayer graphene at the wafer-scale. Alloying in vdW HSs therefore provides a broad range of opportunities for the large-scale growth of tunable moiré materials.

**Methods**

*Two-step growth of $WS_ySe_{2-y}/WS_xSe_{2-x}$*

The two-step CVD growth of $WS_ySe_{2-y}/WS_xSe_{2-x}$ is done on 300 nm $SiO_2/Si$ substrates (2×2 cm$^2$) cleaned by ultrasonication in acetone, isopropanol and DI water for 5 minutes each. For the growth of the first monolayer (sulfur-rich $WS_xSe_{2-x}$), the substrate is placed face-down on a quartz boat containing 15 mg of $WO_{2.9}$ powder (Alfa Aesar 99.99%) and 0.5-1 mg of NaCl.[56] The substrate is positioned at the center of a one zone furnace in a 2-inch quartz tube. 70 mg of sulfur powder (Sigma Aldrich, 99.98%) is placed upstream, 1 cm out of the furnace (~200 °C) and selenium (Beantown Chemical, 99.999%) is provided by the background pressure in the tube. Before growth, two cycles of pumping (down to 10 mTorr) and purging with Ar (99.999%) are done to remove $H_2O$ and $O_2$. The growth is carried out at atmospheric pressure under 200 sccm Ar flow. The temperature is ramped at a rate of 25 °C/min and held at 850 °C for 6 min. The furnace is then cooled down to room temperature using an air pump and the samples are taken out and stored in a glovebox.

In the second step, selenium rich monolayer $WS_ySe_{2-y}$ is grown on $WS_xSe_{2-x}/SiO_2/Si$ using the same method. 100-130 mg $WO_{2.9}$ and 18-27 mg NaCl are used with 200 sccm of 10% $H_2$ in Ar as the carrier gas (99.999%). The sulfur to selenium ratio is controlled by independently changing the temperature and quantity of both precursors. 0-25 mg of sulfur powder and 20-30 mg of selenium powder are used, and their respective temperature is controlled by their position in the furnace (1.25-3 cm outside the furnace for S and 0-0.75 cm inside the furnace for Se). The estimated Se temperature is ~400 °C.

*Transmission electron microscopy*

For TEM characterization, $WS_ySe_{2-y}/WS_xSe_{2-x}$ vdW HSs are transferred onto lacey carbon on Cu mesh grids (Ted Pella) using an etching-free transfer method with a PMMA film[57]. To minimize damage to the HS, ultrasonication at low power (25W, ~10-30s) is used to detach the $PMMA/WS_ySe_{2-y}/WS_xSe_{2-x}$ from the substrate only when needed. TEM is carried out using a FEI Tecnai Osiris microscope operated at 200 kV in TEM bright field mode. Energy dispersive X-ray spectroscopy (EDX) is done in TEM mode. The selenium content is determined using the selenium L$\alpha$ to tungsten M$\alpha$ peaks ratio of $WS_ySe_{2-y}/WS_xSe_{2-x}$. $WSe_2$ powder is used as a reference to calibrate the compositions. The Se content from the bottom monolayer $WS_xSe_{2-x}$ regions is subtracted from the total content to determine the top $WS_ySe_{2-y}$ composition and compute $\Delta Se$. If no $WS_xSe_{2-x}$ regions are present (such as in continuous HS), the Se content of the bottom layer is measured on $WS_xSe_{2-x}$ grown in identical conditions. Due to the small size of the 150 nm nucleus presented in Fig. 3i, electron beam damage prevents the direct measurement of its composition. We therefore determine the composition using slightly larger nuclei (300 nm). Given their similar small size, both nuclei likely formed approximately at the same time, and therefore under the same conditions.

*Photoluminescence*

Micro-PL measurements are done using a 40x (0.6 NA) objective with 532 nm excitation at 77K. Due to the PL quenching observed in as-grown $WS_ySe_{2-y}/WS_xSe_{2-x}$ on $SiO_2/Si$, the vdW HSs were encapsulated with hBN. First, the HS are transferred to clean $SiO_2/Si$ substrates with the same method used for transfer to TEM grids. To facilitate encapsulation, the lateral size of the HS is reduced to ~10 μm by reactive ion etching ($CHF_3/O_2$). After hBN encapsulation, contaminants are squeezed out of the interface using an atomic force microscope tip in contact mode[58].

*Modelling of the energy of nuclei and molecular dynamics simulations*

To better understand the moiré formation mechanisms, we estimate the evolution of the energy *vs.* lattice parameter for triangular $WS_ySe_{2-y}$ nuclei on $WS_xSe_{2-x}$. To keep our analysis general to vdW HSs, and capture the relevant physics, we model the nuclei with a simple monolayer hexagonal 2D lattice with one atom per unit cell. Since the two layers are coupled only by weak vdW interactions, we compute the intralayer and interlayer interaction independently. The intralayer interactions are modelled by a harmonic bond stretching potential $V_{ij} = \frac{1}{2}k(r_{ij} - r_{ij}^0)^2$ and a bond angle potential $V_{ijk}^\theta = \frac{1}{2}k_\theta(\theta_{ijk} - \theta_0)^2$. Here, the $r_{ij}$ are the bond lengths for atoms *i* and *j* and the $\theta_{ijk}$ are the bond angles formed by atoms *i*, *j*, and *k*. The $r_{ij}^0$ and $\theta_0$, are the equilibrium values. We note that the bond angle potential is always zero for the energy calculations but deviates from zero in the molecular dynamics simulations. The force constants $k$ = 141.5 N/m and $k_\theta$ =11 eV rad$^{-2}$ are fitted to reproduce the experimental Young's modulus and Poisson ratio of $WS_2$[42]. We use the same $k$ and $k_\theta$ for all compositions since the differences in the mechanical properties of $WS_2$ and $WSe_2$ are minimal[42], and the alloy properties can be expected to be in the same range[59]. $\theta_0$ is set at $\pi/3$ and $r_{ij}^0$ is the equilibrium lattice parameter of $WS_ySe_{2-y}$. To model the edge stress, we adjust the equilibrium bond length at the edges of the nuclei. A 5% contraction of the equilibrium bond length for the outmost unit cells ($r_{edges}^0$) is used to model tensile stress, which yields edge stresses comparable with *ab initio* calculations[41]. The bulk strain energy $U_{bulk}$ is obtained by summing $V_{ij}$ for all bonds, assuming $r_{edges}^0 = r_{bulk}^0$. The edge strain energy $U_{edges}$ is then given by $U_{edges} = U_{tot} - U_{bulk}$, where $U_{tot}$ is the sum the $V_{ij}$ for all a bonds, assuming the shortened $r_{edges}^0$.

To model interlayer interactions, we assume that the bottom layer remains fixed, and that the interaction energy depends only on the relative local stacking of the two layers. This reduces the complex interactions, such as those resulting from atomic reorganization within the unit cell, to a 2D potential $U(x, y)$ with the periodicity of the bottom layer. The exact nature of $U(x, y)$ depends on the materials system and is not the focus of this work. Nonetheless, the general features of moiré formation can be captured as long as the depth and width of $U(x, y)$ are in the correct range. For simplicity, we therefore choose a potential formed by a gaussian well at each lattice site of the bottom layer: $U(\boldsymbol{r}) = -\sum_{\boldsymbol{R}} U_0 \exp\left[-\frac{1}{2}\left(\frac{\boldsymbol{r}-\boldsymbol{R}}{\sigma}\right)^2\right]$, with $\boldsymbol{R}$ the lattice vectors of the bottom layer and $U_0$ and $\sigma$ the potential well depth and width, respectively. The parameters $U_0$ and $\sigma$ are estimated using a combination of experimental data and *ab initio* calculations results[44]. In fact, our results indicate that $\Delta$Se=17% approaches the minimal lattice mismatch allowing for moiré formation. This implies that $\Delta E_{int} + \Delta U_{bulk} \approx 0$ at this composition. Given the experimental Young's moduli $k$ (168 and 188 N/m) and Poisson ratios $\nu$ (0.19 and 0.22) of WSe$_2$ and WS$_2$[42], and assuming a linear dependence as a function of the composition of the WS$_x$Se$_{2-x}$ alloy[59], the presence of moiré patterns in vdW HSs with $\Delta$Se=17% suggests that $\Delta E_{int} \approx 0.011$ J/m$^2$. The parameter $U_0$ is related to $\Delta E_{int}$ by $U_0 > A \times \Delta E_{int}$, where $A$ is the area of the unit cell. The inequality indicates that some atoms remain in the potential wells, even in the relaxed configuration. It should be noted that the estimation of $\Delta E_{int}$ is highly sensitive to the experimental accuracy of $\Delta$Se. On the other hand *ab initio* calculations find that the corrugation of the interlayer interaction energy for the sliding of bilayer TMDCs is ~0.035 J/m$^2$ [44]. We therefore use a value that yields $E_{int} = 0.02$ J/m$^2$ ($U_0 = 10.7$ meV), which falls inside the range of our experimental estimation of $\Delta E_{int}$ and the *ab initio* predictions. We note that a wide range of $U_0$ leads to qualitatively similar behaviors. Furthermore, we choose $\sigma = a_{WS_2}/9$, where $a_{WS_2}$=3.153 Å is the lattice constant of WS$_2$, to reproduce the *ab initio* calculated width of the interlayer interaction potential wells in TMDCs[44].

For the molecular dynamics simulations presented in the supplementary information, the same model is used, but the interlayer potential is modified to obtain thermally activated relaxation within a reasonable simulation time and number of atoms. Increasing the magnitude of $U_0$ was not found to qualitatively affect the relaxation behavior, other than shifting it to smaller sulfur content. A mass of $M_W + yM_S + (2 - y)M_{Se}$ is used for the atoms of the nuclei. The relaxation of triangular WS$_y$Se$_{2-y}$ flakes on WS$_2$ is simulated using a 5 fs time step with the velocity Verlet algorithm[60]. To prevent concerted motion, the simulation is initiated by linearly ramping down the interlayer potential from 2.5 $U_0$ to $U_0$ in 200 ps.

**Acknowledgements**

M.F.-D. acknowledges support from the Natural Sciences and Engineering Research Council of Canada (NSERC) and Fonds de recherche du Québec (FRQNT). M.F.D. and F.X. acknowledge the partial support from Government of Israel. K.W. and T.T. acknowledge support from JSPS KAKENHI (Grant Numbers 19H05790, 20H00354 and 21H05233).

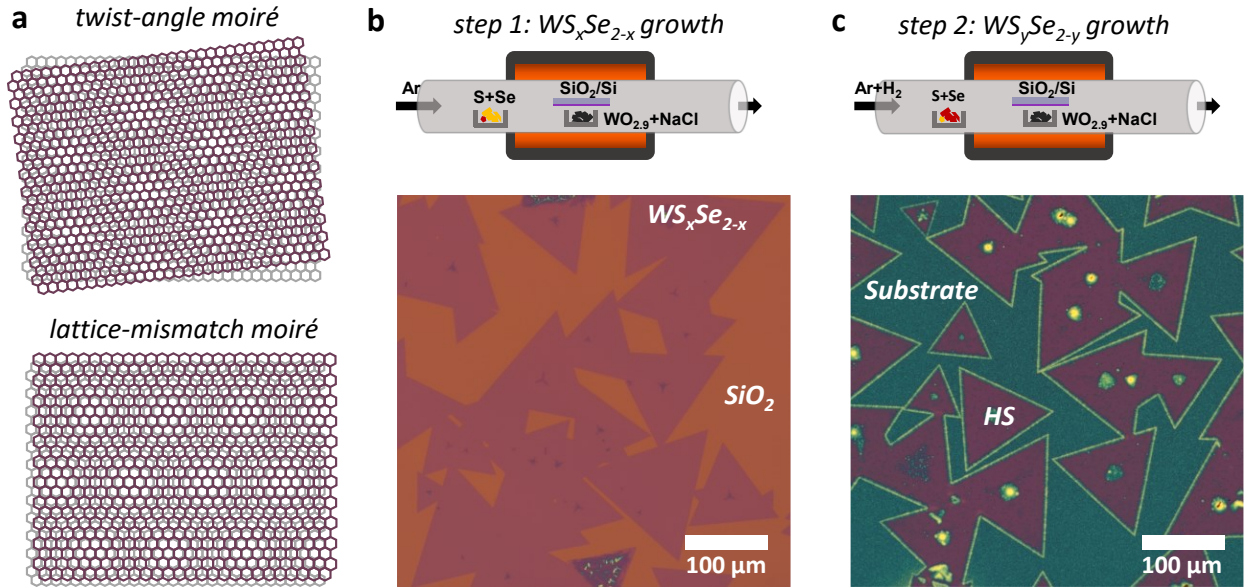

**Figure 1. Growth of moiré vdW heterostructures. a**, Schematics of moirés resulting from the twist between two 2D layers (top) and the stacking of lattice-mismatched layers (bottom). The twist angle and the magnitude of lattice mismatch determine the moiré periods in these two cases, respectively. **b**, First step of the two-step CVD growth process of $WS_ySe_{2-y}/WS_xSe_{2-x}$. During this step, monolayer $WS_xSe_{2-x}$ flakes grown on $SiO_2/Si$, as seen in the optical micrograph in the bottom. **c**, Second step of the vdW HSs growth process. In this step, $WS_ySe_{2-y}$ is grown on $WS_xSe_{2-x}$. The darker purple triangular flakes in the optical micrograph are the $WS_ySe_{2-y}/WS_xSe_{2-x}$ vdW HS. The teal regions are multilayer $WS_ySe_{2-y}$ on $SiO_2$.

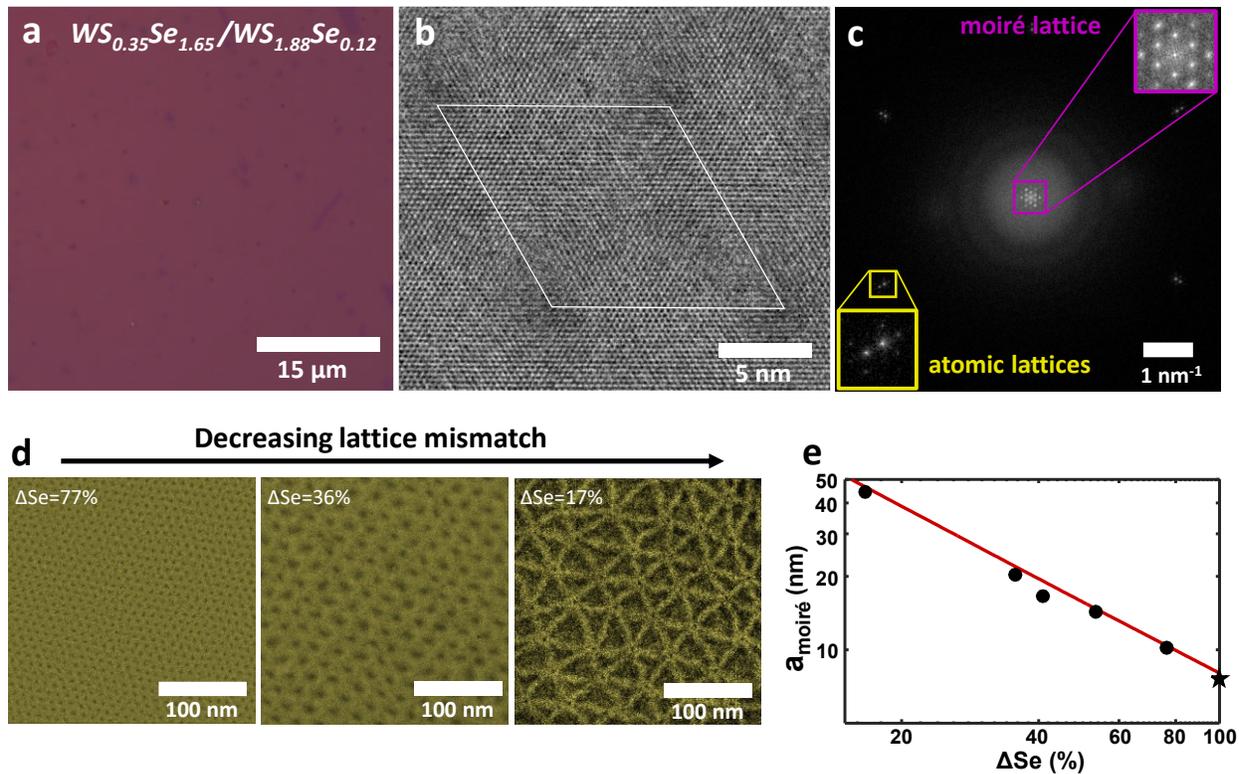

**Figure 2. Tunable moiré pattern in $WS_ySe_{2-y}/WS_xSe_{2-x}$ vdW HSs. a,** Optical micrograph of a $WS_{0.35}Se_{1.65}/WS_{1.88}Se_{0.12}$ vdW HS. The composition is determined by EDX. **b,** High magnification TEM micrograph of the moiré unit cell of the $WS_{0.35}Se_{1.65}/WS_{1.88}Se_{0.12}$ sample in (**a**). **c**, FFT of a large area TEM image of the sample shown in (**a, b**). The yellow and magenta squares show the atomic and moiré reciprocal lattices, respectively. Enlarged images of these regions are shown in the bottom left and top right. **d**, False colored TEM micrographs of the moirés of relaxed vdW HSs with different compositions and moiré periodicity. The moiré pattern observed for ΔSe=17% (dark triangles) likely result from atomic reconstructions that create domain boundaries, similar to those seen in small angle twisted bilayers[40]. **e**, Composition dependance of the moiré lattice constant. The filled circles are samples grown in this work and the star is the measured moiré period of $WSe_2/WS_2$ from Ref. [16].

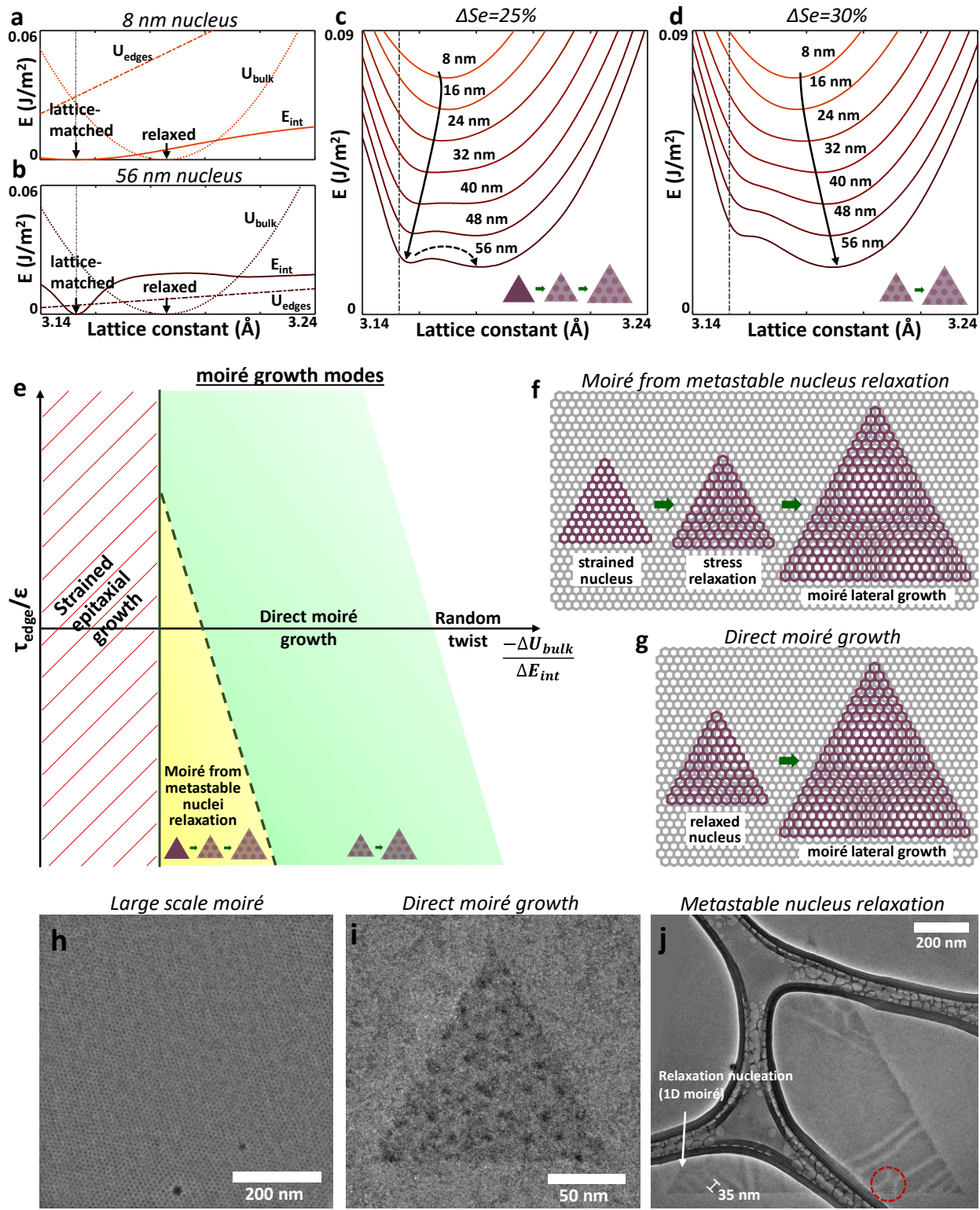

**Figure 3 Moiré growth mechanisms (a, b)** Simulated lattice parameter dependence of the edge strain energy, bulk strain energy and interlayer interaction energy for triangular WS$_{1.5}$Se$_{0.5}$ nuclei on WS$_2$. **a** $L$=8 nm, **b** $L$=56 nm (details in Methods). The black arrows indicate the lattice-matched

and relaxed lattice parameters. **(c, d)** Energy vs. lattice parameter of the top layer for triangular nuclei of increasing $L$. **c,** $WS_{0.5}Se_{1.5}/WS_2$, **d,** $WS_{0.6}Se_{1.4}/WS_2$. The vertical dashed lines indicate the lattice parameter of the $WS_2$ bottom layer. The black arrow illustrates the path followed by growing nuclei to minimize their energy. The dashed arrow shows the transition from the metastable lattice-matched state to the relaxed state. **e**, Schematic of the moiré growth modes as a function of the edge stress ($\tau_{edge}$), bulk strain ($\varepsilon$), bulk strain energy, and interlayer interaction energy. **f,** Schematic of the moiré formation process from metastable strained nuclei. This growth mode corresponds to the yellow region in (**e**). **g,** Schematic of the direct moiré formation mechanism. This growth mode corresponds to the green region in (**e**). **h**, Large-scale TEM micrograph of relaxed continuous $WS_{0.35}Se_{1.65}/WS_{1.88}Se_{0.12}$ showing a homogeneous moiré pattern, likely formed by the direct moiré growth mode. **i,** TEM micrograph of a 150 nm $WS_{1.12}Se_{0.88}$ nucleus on $WS_{1.88}Se_{0.12}$. **j**, TEM micrograph of a $WS_{1.42}Se_{0.58}$ nucleus on $WS_{1.75}Se_{0.25}$ after uniaxial strain relaxation nucleation at its corners. The white arrow indicates uniaxial stress relaxation nucleating at corners, evidenced by parallel 1D moiré stripes with a periodicity of ~ 35 nm. The red circle shows the beginning of biaxial strain relaxation at the edges. The brightness and contrast are locally adjusted to reduce the visibility of artefacts and e-beam damage spots.

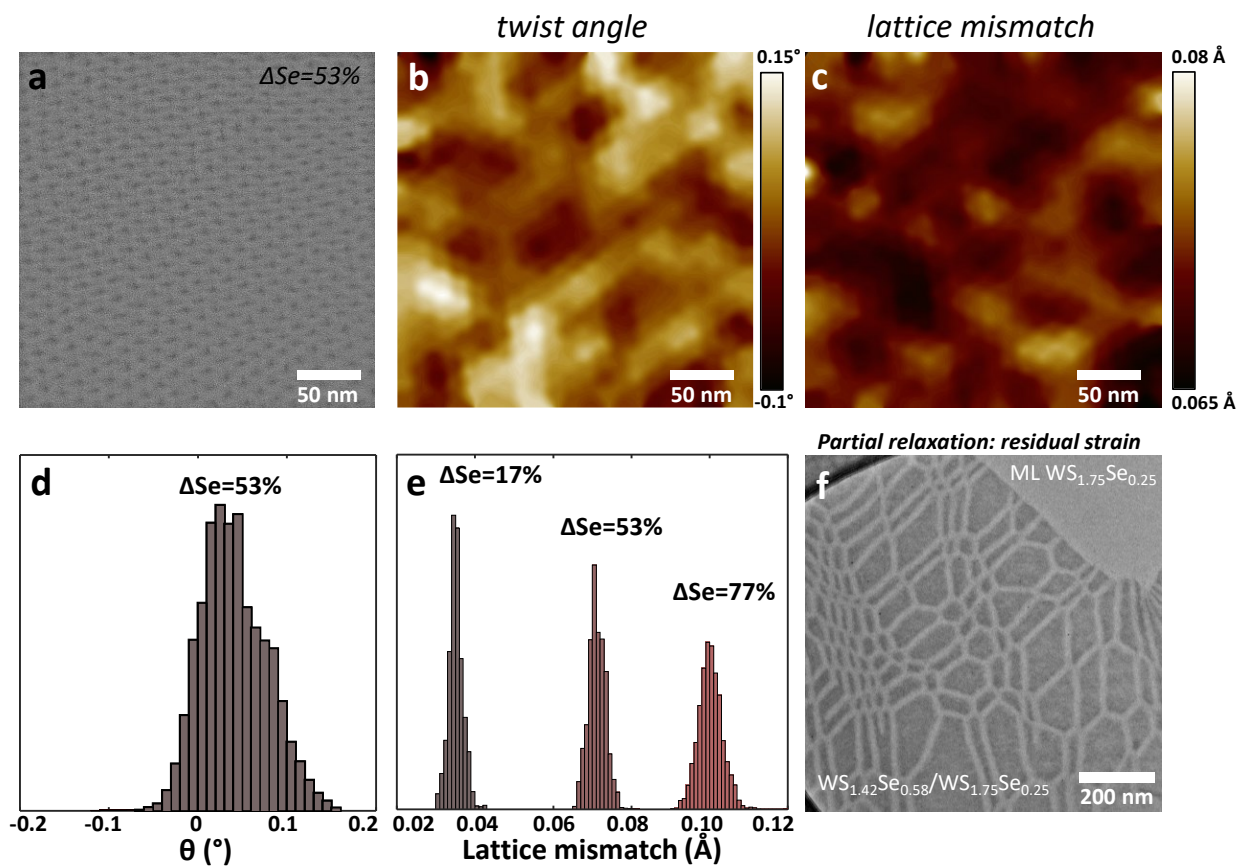

**Figure 4 Moiré homogeneity a**, TEM micrograph of the moiré of $WS_{0.82}Se_{1.18}/WS_{1.88}Se_{0.12}$ (ΔSe=53%). **b**, Map of the local twist angle for the image in (**a**) calculated using the fitted moiré lattice vectors at each lattice site. **c**, Map of the local lattice mismatch for the image in (**a**). **d**, Histogram of the local twist angle for the image in (**a**, **b**). **e,** Histograms of the local lattice mismatch for samples with various compositions. **f**, TEM micrograph of a partially relaxed $WS_{1.42}Se_{0.58}/WS_{1.75}Se_{0.25}$ flake showing large residual strain.

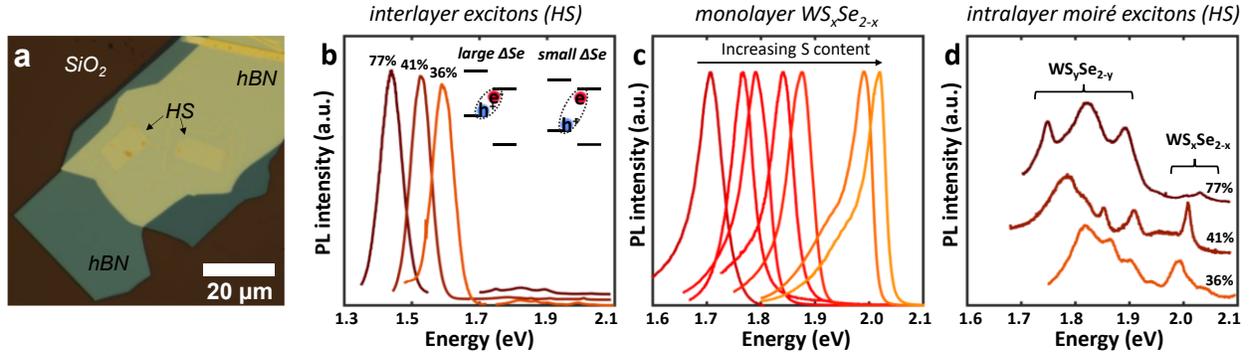

**Figure 5 Excitons in grown moiré vdW HSs a**, Optical micrograph of $WS_{0.14}Se_{1.86}/WS_{1.87}Se_{0.13}$ encapsulated with exfoliated hBN flakes. **b**, Photoluminescence (PL) spectra (T=77K) of three $WS_ySe_{2-y}/WS_xSe_{2-x}$ vdW HSs (ΔSe=77% (brown), ΔSe=41% (dark red) and ΔSe=36% (orange)). The spectra are dominated by the IX (between 1.6 and 1.3 eV). **c**, PL spectra of monolayer $WS_xSe_{2-x}$ flakes with various compositions grown on $SiO_2/Si$. **d**, Close-up of the PL spectra in (**b**) showing the intralayer moiré excitons of the $WS_ySe_{2-y}$ Se-rich top layer (1.75-1.91 eV). No clear splitting of the intralayer exciton peak of the $WS_xSe_{2-x}$ S-rich bottom layer is observed (1.99-2.04 eV).

# Supplementary Information for

# Scalable van der Waals epitaxy of tunable moiré heterostructures

Matthieu Fortin-Deschênes[1*], Kenji Watanabe[2], Takashi Taniguchi[3], Fengnian Xia[1*]

1. Department of Electrical Engineering, Yale University, New Haven, Connecticut 06511, United States
2. Research Center for Functional Materials, National Institute for Materials Science, 1-1 Namiki, Tsukuba 305-0044, Japan
3. International Center for Materials Nanoarchitectonics, National Institute for Materials Science, 1-1 Namiki, Tsukuba 305-0044, Japan

Email: matthieu.fortin-deschenes@yale.edu, fengnian.xia@yale.edu

# Contents

## 1. Growth of monolayer $WS_xSe_{2-x}$

Sulfur rich monolayer $WS_xSe_{2-x}$ was grown on 300 nm $SiO_2$/Si (details in Methods). Most grown $WS_xSe_{2-x}$ flakes are triangular monolayers with lateral dimensions between 50-200 μm (Fig. S1a). Two types of flakes were found: flat monolayer (Fig. S1b) and monolayers with a central 3D nucleus (Fig. S1c). The formation of the 3D nuclei in Fig. S1c may be due to the presence of pre-existing defects on the $SiO_2$ surface, whereas the flat flakes in Fig. S1b may form by homogeneous nucleation on clean $SiO_2$.

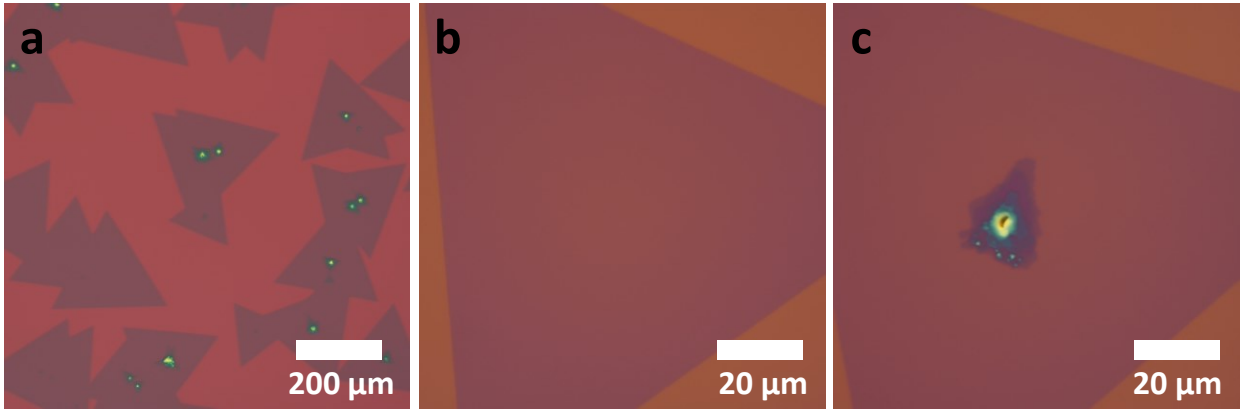

**Figure S1. Monolayer $WS_xSe_{2-x}$ on $SiO_2$. a**, Optical micrograph of $WS_xSe_{2-x}$ monolayers grown on 300nm $SiO_2$/Si. **b**, flat triangular monolayer $WS_xSe_{2-x}$ flake. **c**, monolayer $WS_xSe_{2-x}$ flake with central 3D nucleus. The approximate composition, based on the growth parameters, is $WS_{1.9}Se_{0.1}$.

## 2. Growth mode and orientation of the top WS$_y$Se$_{2-y}$ layer

The vdW HSs grown in this work formed by the coalescence of WS$_y$Se$_{2-y}$ domains with lateral dimensions of up to ~10 μm on single crystal WS$_x$Se$_{2-x}$ flakes (Fig. S2a). Together with the TEM data presented in Fig. 2, Fig. 3 and Fig. 4, the optical micrograph in Fig. S2a indicates that the lattices of WS$_y$Se$_{2-y}$ and WS$_x$Se$_{2-x}$ are rotationally aligned. In fact, the partial growth experiment shows that the edges of the top and bottom layers are aligned. Furthermore, both 2H and 3R stacked domains are present, as indicated by the presence of two orientations of triangular domains, rotated by 180° from each other (Fig. S2a). Moreover, a layer-by-layer growth mode was observed, as shown in Fig. S2b, where some most of the WS$_x$Se$_{2-x}$ flake is covered by monolayer WS$_y$Se$_{2-y}$ and some small bilayer WS$_y$Se$_{2-y}$ nuclei are beginning to grow.

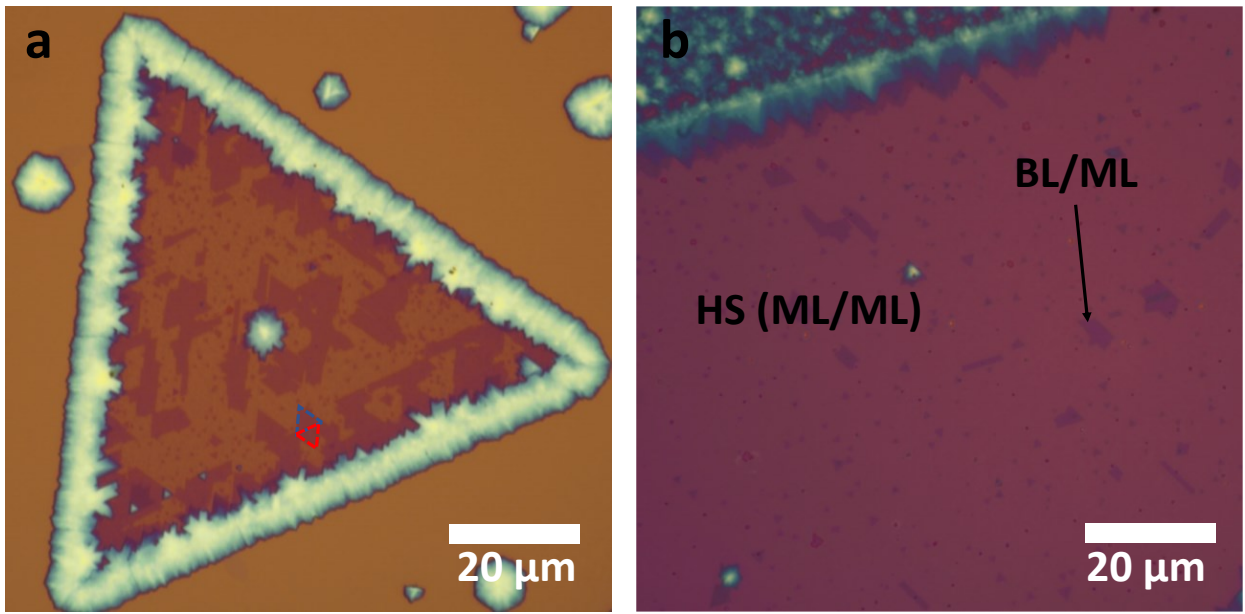

**Figure S2. Growth modes of top WS$_y$Se$_{2-y}$ layer. a**, Partial growth of WS$_y$Se$_{2-y}$/WS$_x$Se$_{2-x}$ showing a mix of 2H and 3R WS$_y$Se$_{2-y}$ domains on the single crystal WS$_x$Se$_{2-x}$ triangular flake. The two types of stackings are evidenced by the presence of triangular top WS$_y$Se$_{2-y}$ domains rotated by 180° from each other (such as the red and blue triangles). **b**, Continuous WS$_y$Se$_{2-y}$/WS$_x$Se$_{2-x}$ vdW HS with small bilayer (BL) WS$_y$Se$_{2-y}$ on WS$_x$Se$_{2-x}$ regions.

The nucleation of a single orientation is important for the scalability of moiré vdW HSs growth. While not studied in detail in this work, we found that single orientation nucleation of $WS_xSe_{2-x}$ with >99.5% alignment can occur under some growth conditions, as shown in Fig. S3 (Growth parameters: 120 mg $WO_{2.9}$, 22mg NaCl, 20 mg Se powder, 20 mg S powder, 850 °C (1 min), 25 °C/min ramp, 200 sccm (10% $H_2$ in Ar). In fact, triangular bilayer $WS_xSe_{2-x}$ domains here were formed with perfect alignment. This shows that nucleation can be controlled to grow single-crystal TMDCs vdW HSs, as mentioned in the main text.

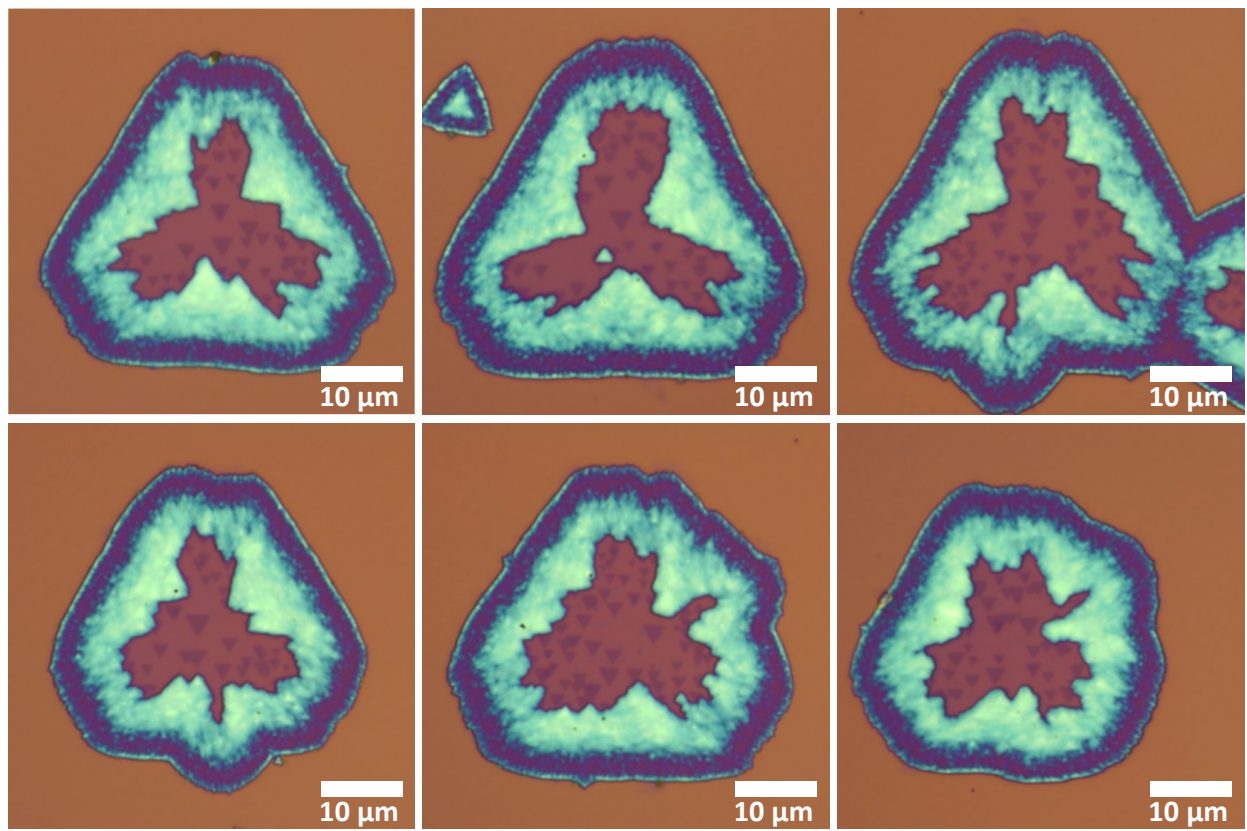

**Figure S3. Single orientation nucleation of bilayer $WS_xSe_{2-x}$.** Optical micrographs of $WS_xSe_{2-x}$ flakes. Only one orientation of the top layer was observed, as illustrated by the orientation of the dark purple (bilayer) triangles in the middle of the flakes. Growth parameters: 120 mg $WO_{2.9}$, 22mg NaCl, 20 mg Se powder, 20 mg S powder, 850 °C (1 min), 25 °C/min ramp, 200 sccm (10% $H_2$ in Ar). The approximate composition, based on the growth parameters, is $WS_{1.3}Se_{0.7}$.

## 3. TEM diffraction characterization

TEM diffraction experiments further confirm the rotational alignment and stress relaxation in the grown vdW HSs, as shown in Fig. S4. Moreover, the diffracted intensity from $WS_xSe_{2-x}$ and $WS_ySe_{2-y}$ is similar, which implies that both layers have the same thickness (i.e., monolayers), in agreement with the optical microscopy contrast.

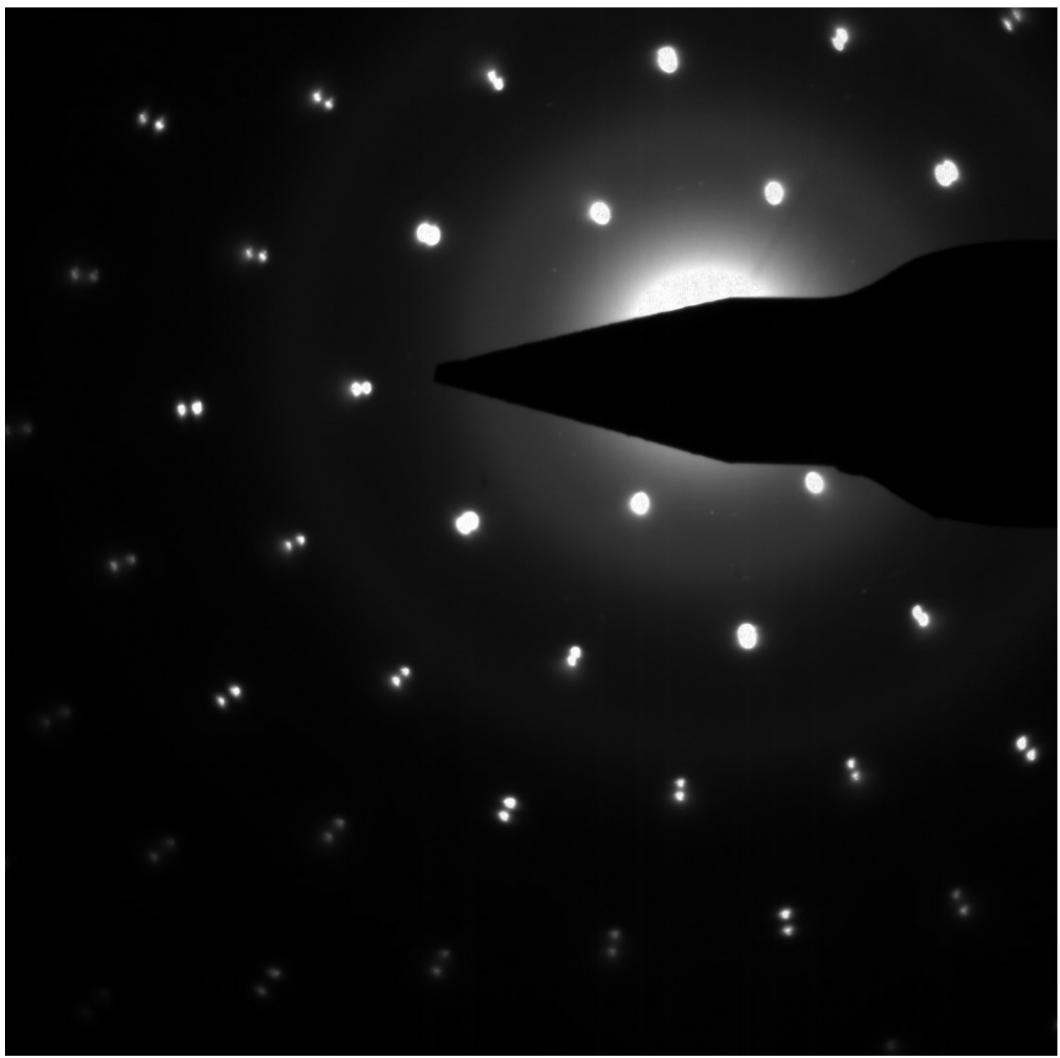

**Figure S4. TEM diffraction pattern of $WS_{0.35}Se_{1.65}/WS_{1.88}Se_{0.12}$.**

## 4. Energy-dispersive X-ray spectra (EDX)

As described in the methods section, the composition of the vdW HSs is determined by Energy-dispersive X-ray spectroscopy (EDX). The EDX spectra of monolayer $WS_{1.88}Se_{0.12}$ (bottom layer), and of $WS_ySe_{2-y}/WS_xSe_{2-x}$ vdW HSs with $\Delta Se$ between 17-77% are shown in Fig. S5 below.

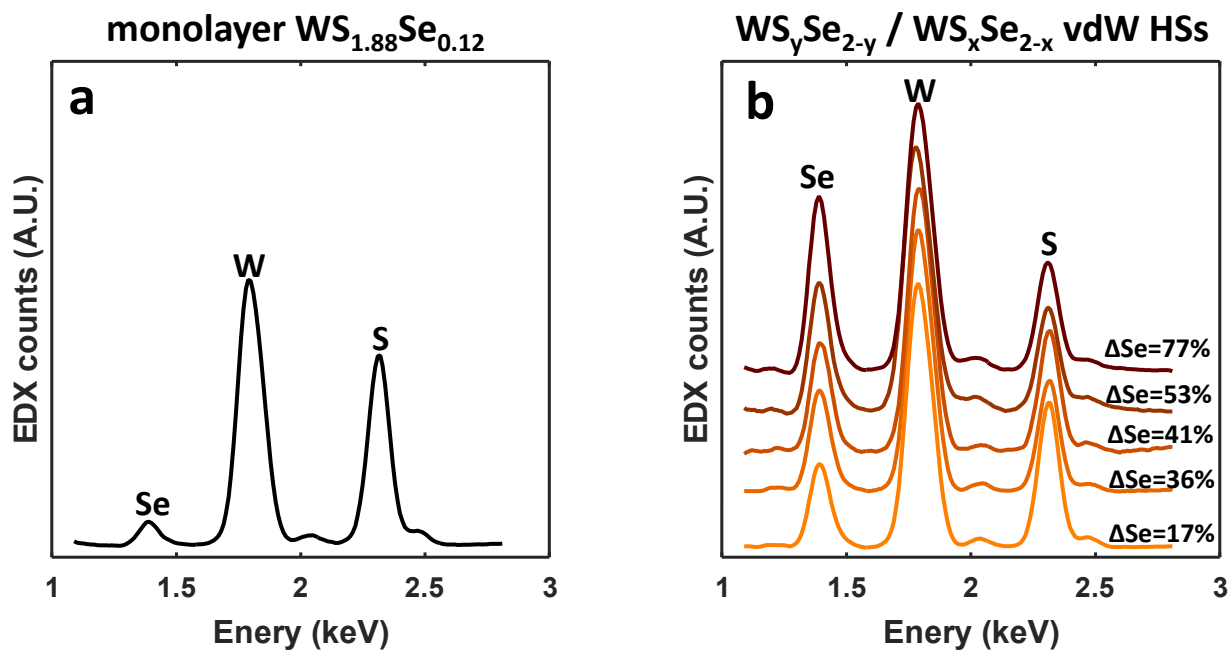

**Figure S5 Energy-dispersive X-ray spectra (EDX) the grown materials. a**, EDX spectrum of monolayer $WS_{1.88}Se_{0.12}$ grown in the same conditions used for the bottom $WS_xSe_{2-x}$ monolayers. **b**, EDX spectra of $WS_ySe_{2-y}/WS_xSe_{2-x}$ vdW HSs with $\Delta Se$ between 17-77%.

## 5. Energy of small nuclei

To better understand the moiré growth modes, we estimated the strain and interlayer interaction energies as a function of the lattice parameter for nuclei of different sizes and compositions. The energy is calculated using the model described in the methods section for triangular nuclei such as those presented in Fig. S6. As mentioned in the main text, the goal of the simulations is to establish the relevant physics and mechanisms of moiré formation during growth, general to rotationally aligned vdW HSs. As a result, here we utilize a simplified model of a single layer hexagonal lattice in a 2D periodic potential to represent the more complex atomic structure of a single layer $WS_ySe_{2-y}$ nuclei on $WS_xSe_{2-x}$. This simple model highlights the role of interlayer interactions, edge stress and bulk strain in shaping the moiré growth modes.

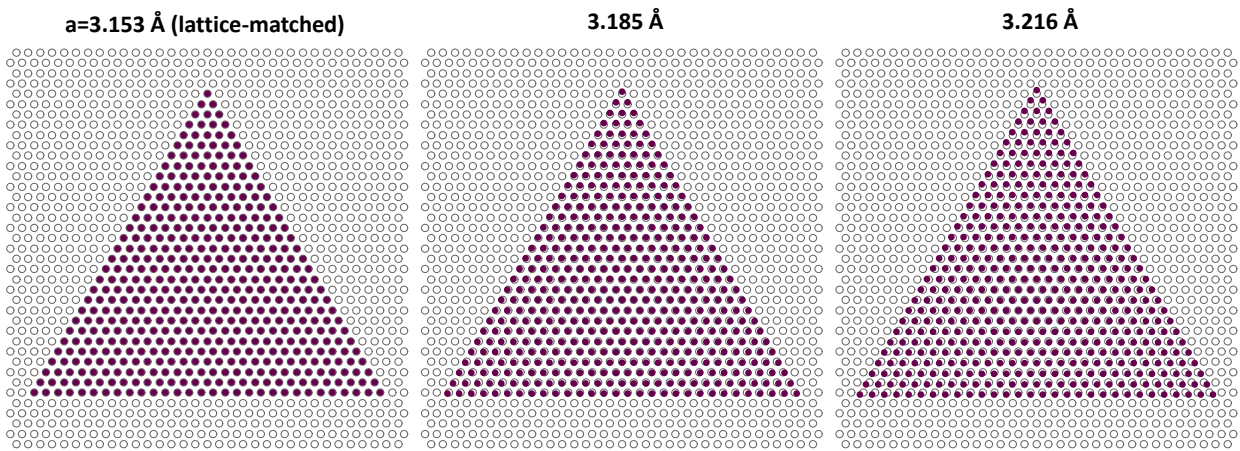

**Figure S6. Model of triangular nuclei.** The top layer triangular nuclei are modelled as a hexagonal lattice (purple circles) with each atom bonded to its nearest neighbors, as described in the methods section. The gray hollow circles indicate the fixed lattice sites of the bottom layer, which are used to define the interlayer interaction potential with a gaussian well at each lattice site. To model tensile (compressive) edge stress, the equilibrium interatomic distance is shortened (increased) for the outmost bonds. The total energy is then computed for nuclei with different lattice parameters.

As explained in the main text, when the bulk and edge stress of the nuclei have opposite signs (such as in $WS_ySe_{2-y}/WS_xSe_{2-x}$ vdW HSs where $y < x$), the edge stress can prevent bulk stress relaxation. As shown in Fig. 3c and 3d and Fig. S7, two minima form in the potential energy surface, as the nuclei grow. If the edge stress is sufficient, as compared to the bulk stress, it can drive small nuclei in the lattice-matched minimum. However, if the lattice mismatch is large enough, the relaxed configuration becomes the most stable state in larger nuclei, and stress relaxation can occur.

Interestingly, growing nuclei can be trapped in the metastable state even without edge stress, as shown in Fig. S7. This occurs at very small lattice mismatch, near the threshold for moiré formation. In fact, the minimum of the interlayer interaction energy *vs* nucleus lattice constant ($E_{int}$ ($a_{nucleus}$)) is broad in small nuclei. Partial stress relaxation can therefore occur while keeping most atoms of the nuclei near the lattice-matched sites. As the nuclei grow, $E_{int}$ ($a_{nucleus}$) narrows and the nuclei naturally relax into the lattice-matched configuration if the bulk strain energy is small enough. This effect can be visualized for ΔSe between 21-23% in Fig. S7, when there is no edge stress (identical interatomic distances at the edges and in the bulk for our model ($r^0_{edge} = r^0_{bulk}$)). Furthermore, we note that as the edge stress becomes larger (such as for $r^0_{edge} = 0.95\ r^0_{bulk}$ and $r^0_{edge} = 0.9\ r^0_{bulk}$ in Fig. S7), metastable lattice-matched nuclei form at larger ΔSe. On the other hand, if the edge stress has an opposite sign, direct moiré growth is favored in nuclei with even smaller lattice mismatch.

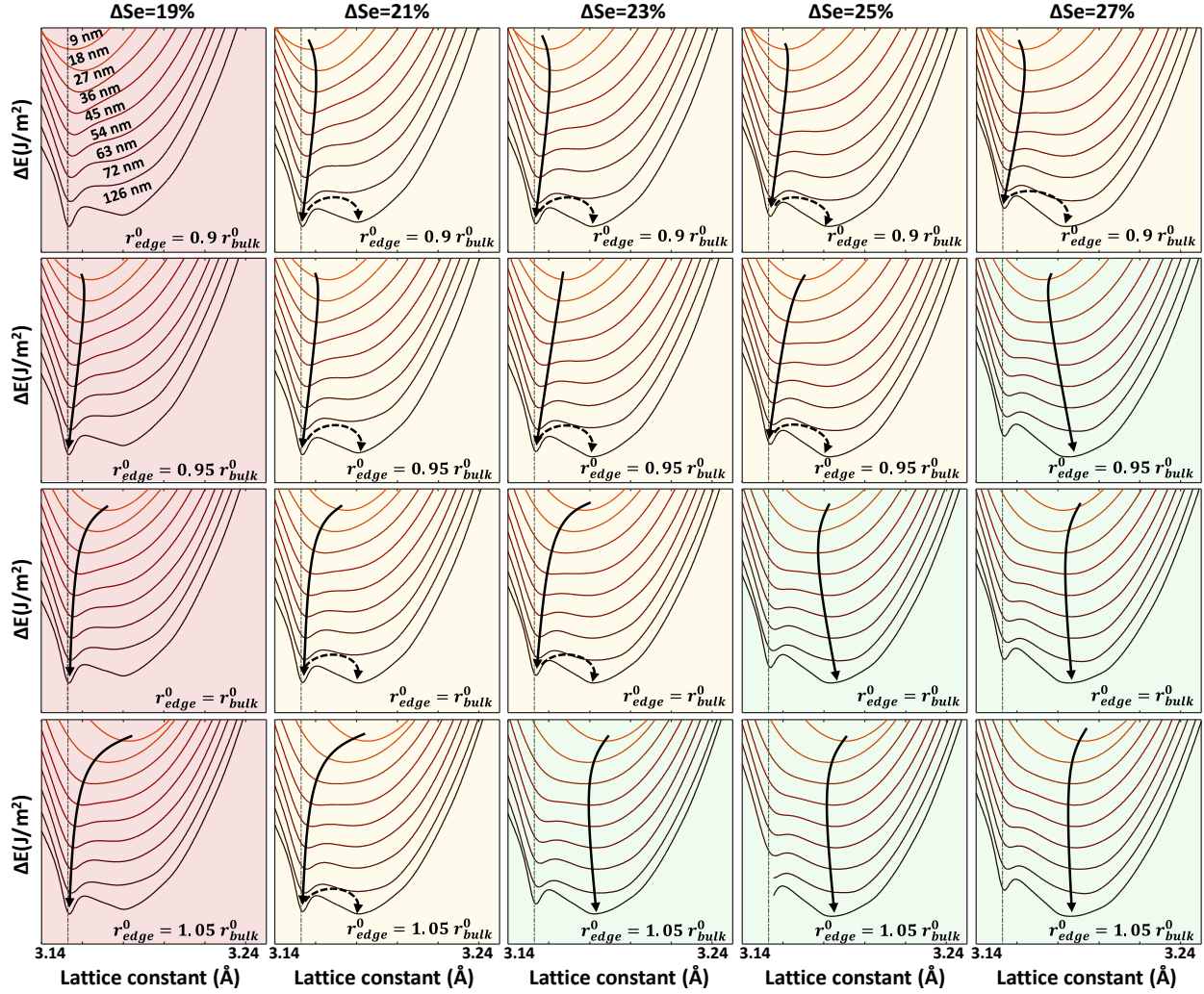

**Figure S7. Energy *vs.* lattice constant for growing nuclei.** Calculated energy *vs.* lattice constant for triangular nuclei of increasing lateral size. The edge lengths used in this calculation are the same in all graphs and are shown in the top left. The lattice mismatch (proportional to ΔSe) increases from left to right. The edge stress (modelled by changing the equilibrium interatomic distance at the edges) goes from highly tensile to compressive, when going from the top to the bottom. path followed by growing nuclei to minimize their energy is indicated by the black arrows. The dashed arrows indicate the thermally activated relaxation to form moiré patterns. The color of the background indicates the favored moiré growth mode. Red: epitaxial growth without moiré formation. Yellow: formation of lattice-matched metastable nuclei followed by moiré formation. Green: direct moiré growth.

## 6. Molecular dynamics simulations

To better understand the stress relaxation mechanisms of metastable lattice-matched nuclei, we carried out molecular dynamics (MD) simulations of the relaxation with the same model used for calculation of the nuclei's energy *vs.* lattice constant, using Matlab. Since the metastable state only forms at small lattice mismatch (long period moiré), a large number of atoms is required for the nuclei to contain several moiré unit cells. We therefore increase the interlayer interaction potential to shift the metastable nuclei relaxation to larger lattice mismatch to reduce the simulation time. We use $U_0$ = 48 meV and $\sigma = a_{WS_2}/12$, as well as $r_{edge}^0 = 0.9\, r_{bulk}^0$. This allows to visualize several moiré unit cells in nuclei with a smaller number of atoms (~10,000). We note that increasing the magnitude of $U_0$ was not found to qualitatively affect the relaxation behavior, other than shifting it to smaller sulfur content. More details about the simulations can be found in the methods section.

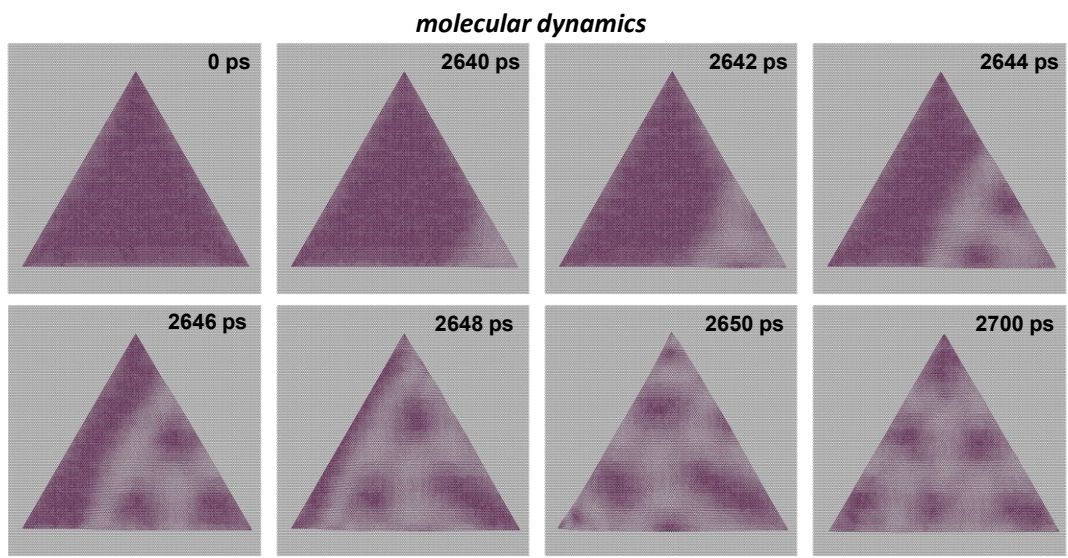

**Figure S8** Molecular dynamics simulation snapshots of strain relaxation for ΔSe=60%. The triangular nucleus has 40.7 nm sides and the temperature before relaxation is 810 °C (constant energy simulation). The nucleus is trapped in the metastable lattice-matched configuration in the first 2640 ps of the simulation. The relaxation nucleates at bottom right the corner by a random event and then rapidly spreads inside the bulk.

The simulations confirm that nuclei can remain trapped in a metastable state and eventually relax by a thermally activated process, as shown in Fig. S8. We note that when ΔSe is too large, the nuclei immediately relax since no energy barrier separates the lattice-matched and relaxed configurations. However, when the nuclei are trapped in the metastable state, stress relaxation nucleates at the corners of the triangles and spreads inside the bulk.

## 7. Calculations of moiré homogeneity

The homogeneity of the twist angle and lattice parameter is calculated from TEM images by taking advantage of the moiré amplification[1]:

$$\vec{k}^i_{moiré} = \vec{k}^i_{WS_xSe_{2-x}} - \vec{k}^i_{WS_ySe_{2-y}} \tag{S1}$$

where the $\vec{k}^i$ are the reciprocal lattice vectors of the moiré and atomic lattices of the two layers, and $i = (1, 2, 3)$ are the labels of the three in-plane lattice vectors. The analysis is carried out on a single TEM image for each sample at a magnification where the atomic lattices are visible (at 64 kX). The $\vec{k}^i_{WS_xSe_{2-x}}$ are measured directly in the FFT of the TEM image. To correct for astigmatism, processing and analysis is carried out in the uncorrected images, but the extracted values are rescaled in the following way. First an ellipse is fitted[2] on the six (100) reciprocal lattice points of the bottom layer in the FFT (principal axes $a$ and $b$ tilted from the $x$ and $y$ axes by $\phi$). The real space and reciprocal space coordinates are then corrected by first rotating the image to align the principal axes to the $x$ and $y$ axes, then rescaling the $(x, y)$ coordinates to make the ellipse circular and rotating the image to its initial orientation. These operations, for the real and reciprocal space, can be written in matrix form to determine the corrected coordinates $(x', y')$ and corrected reciprocal space coordinates $(k_x', k_y')$:

$$\begin{bmatrix} x' \\ y' \end{bmatrix} = \begin{bmatrix} \alpha^{-1}\cos^2\phi + \alpha \sin^2\phi & \sin\phi\cos\phi(\alpha + \alpha^{-1}) \\ \sin\phi\cos\phi(\alpha + \alpha^{-1}) & \alpha^{-1}\sin^2\phi + \alpha\cos^2\phi \end{bmatrix} \begin{bmatrix} x \\ y \end{bmatrix} \tag{S2}$$

and:

$$\begin{bmatrix} k_x' \\ k_y' \end{bmatrix} = \begin{bmatrix} \alpha\cos^2\phi + \alpha^{-1}\sin^2\phi & \sin\phi\cos\phi(\alpha + \alpha^{-1}) \\ \sin\phi\cos\phi(\alpha + \alpha^{-1}) & \alpha\sin^2\phi + \alpha^{-1}\cos^2\phi \end{bmatrix} \begin{bmatrix} k_x \\ k_y \end{bmatrix} \tag{S3}$$

with a scale factor $\alpha = (b/a)^{1/2}$.

$\vec{k}^i_{moiré}$ are then measured along three directions for each moiré lattice sites using the following procedure. First, the contrast is improved by filtering the TEM images using a Hann windows of width $\approx \vec{k}_{moiré}$ centered on the moiré reciprocal lattice points, using the software WSxM [3]. The approximate position of each moiré lattice site is then determined using the particles analysis tool in the ImageJ software[4]. The exact position of each site is then fitted in the filtered image using a 2D gaussian function[5]. The real space moiré lattice vectors are then directly calculated using the fitted (and corrected for astigmatism) coordinates. To obtain the lattice vectors at every pixel of the TEM images, we take the convolution of the calculated values at each site (Dirac delta) with a Hann window function of half-width $a_{moiré}$ where $a_{moiré}$ is the average lattice parameter measured in the TEM image.

Next, we calculate the $\vec{k}^i_{moiré}$ from the fitted real space lattice vectors and determine the $\vec{k}^i_{WS_ySe_{2-y}}$ at every pixel using equation S1 and the $\vec{k}^i_{WS_xSe_{2-x}}$ measured in the FFT and corrected for astigmatism. We then plot the twist angle in Fig. 4b, as the average angle between the three reciprocal lattice vectors of the top and bottom layers at every pixel. Similarly, lattice mismatch in Fig. 4c is the average of lattice parameters difference between the top and bottom layer for the three lattice vectors. The histograms in Fig. 4d and 4e are directly computed from the twist angle and lattice mismatch maps such as those shown in Fig. 4b and 4c.

We note that for the long period moiré in the ΔSe=17% sample, we used a lower magnification image to have a large enough number of moiré unit cells. In the low magnification image (at 26 kX), the atomic lattices are not visible. We therefore determined the direction of $\vec{k}_{WS_xSe_{2-x}}$ using the real-space images, by identifying the orientation of crystal edges. While this leads to larger inaccuracy in the absolute orientation of the computed $\vec{k}_{WS_ySe_{2-y}}$, it is still highly accurate for the determination of the relative twist angle and lattice mismatch due to the large moiré amplification at such a small lattice mismatch.